# Integrated Avalanche Photodetectors for Visible Light

**Salih Yanikgonul,**[1,4] **Victor Leong,**[1,†] **Jun Rong Ong,**[2,*] **Ting Hu,**[3] **Ching Eng Png,**[2] **and Leonid Krivitsky** [1]

[1]*Institute of Materials Research and Engineering, Agency for Science, Technology and Research (A*STAR), 138634 Singapore*
[2]*Institute of High Performance Computing, Agency for Science, Technology and Research (A*STAR), 138632 Singapore*
[3]*Institute of Microelectronics, Agency for Science, Technology and Research (A*STAR), 138634 Singapore*
[4]*School of Electrical and Electronic Engineering, Nanyang Technological University, 639798 Singapore*
[†]*victor_leong@imre.a-star.edu.sg*
[*]*ongjr@ihpc.a-star.edu.sg*

**Abstract:** Integrated photodetectors are essential components of scalable photonics platforms for quantum and classical applications. However, most efforts in the development of such devices to date have been focused on infrared telecommunications wavelengths. Here, we report the first monolithically integrated avalanche photodetector (APD) for visible light. Our devices are based on a doped silicon rib waveguide with a novel end-fire input coupling to a silicon nitride waveguide. We demonstrate a high gain-bandwidth product of $216 \pm 12$ GHz at 20 V reverse bias measured for 685 nm input light, with a low dark current of 0.12 µA. This performance is very competitive when benchmarked against other integrated APDs operating in the infrared range. With CMOS-compatible fabrication and integrability with silicon nitride platforms, our devices are attractive for visible-light photonics applications in sensing and communications.

## Introduction

Integrated photonics platforms are well poised to meet the growing demands of both classical and quantum applications [1, 2]. These platforms can accommodate multiple components on the same chip, including light sources, modulators, and photodetectors [3]. The use of mature CMOS fabrication processes offers scalable manufacturing and deployment of these devices.

On-chip avalanche photodetectors (APDs) are indispensable components of a fully integrated photonics platform. They provide fast detection speeds, high sensitivity down to single-photon levels, and are compatible with waveguide-based designs. The majority of recent research in this area has been geared towards applications in optical communications networks, focusing on operation at infrared telecommunications wavelengths. These devices have been developed on a variety of material platforms, including III-V semiconductors [4], germanium (Ge) [5–7], and Si [8–11].

However, integrated APDs for visible light detection have yet to be demonstrated. These devices will greatly benefit numerous application areas, including visible light communications [12–14], LIDAR [15, 16], biomedical imaging [17, 18], and molecular sensing [19]. They are advantageous for developing scalable systems for quantum information processing as they do not require cryogenic environments, unlike integrated superconducting photodetectors. Integrated APDs also enable the photonic integration of various quantum systems operating at visible wavelengths, such as trapped ions, color centers centers in diamond, quantum dots, and 2D materials [20].

Despite the ubiquity and high performance of free-space APDs for visible light detection, integrated versions present a host of technical challenges. To date, the shortest operating wavelength among integrated APDs is 850 nm, as demonstrated in devices developed for short-reach data communications [9, 11]. To the best of our knowledge, there are no reports of visible-light integrated APDs in the literature. One major difficulty lies in the coupling of input light. Conventional integrated APDs rely on an interlayer transition from an input waveguide above or below the APD [21–23]. However, using the same approach for visible wavelengths would lead to deteriorations in noise and bandwidth performance, due to the much longer coupling length required to achieve efficient coupling.

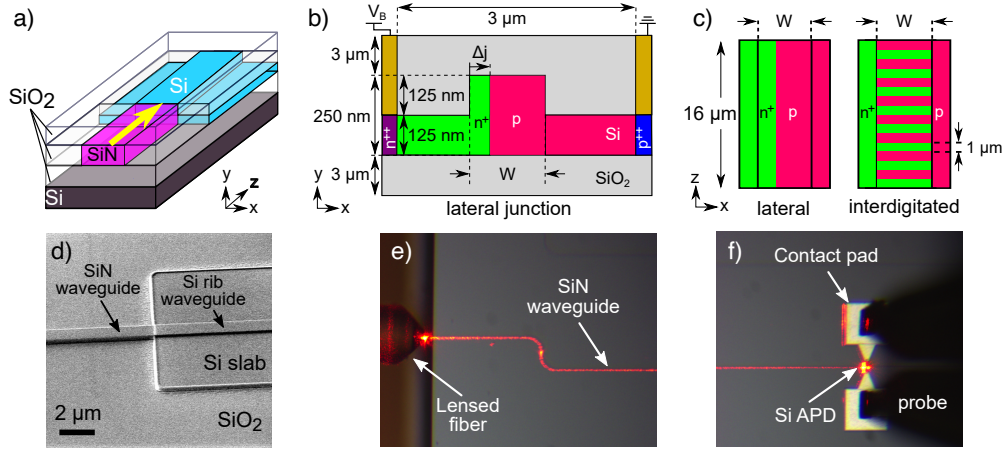

Fig. 1. Device structure and doping configurations. (a) Schematic of the APD device, consisting of a Si rib waveguide end-fire coupled to an input SiN waveguide. The yellow arrow denotes the propagation direction of input light. (b) Cross-sectional view of the Si rib waveguide with a lateral doping profile. The junction placed at a distance $\Delta j$ from the left edge of the waveguide core with a width $W$. A reverse bias voltage $V_B$ is applied via metal contacts deposited on top of heavily-doped $p^{++}$ and $n^{++}$ regions. (c) Top view of the Si rib waveguide, showing the lateral and interdigitated doping profiles. Figures (a)-(c) are not drawn to scale. (d) Scanning electron microscope (SEM) image of a fabricated device without the top $SiO_2$ cladding and metal contacts. (e),(f) Fabricated devices imaged under an optical microscope, showing the lensed fiber coupling and Si APD regions, respectively. The red glow is due to the scattering of the 685 nm input light.

Here, we present the first demonstration of waveguide-coupled APDs for visible light detection. Our devices are fabricated with CMOS-compatible materials, using silicon nitride ($Si_3N_4$, hereafter denoted as SiN) on a silicon-on-insulator (SOI) platform. The device structure is based on a doped Si rib waveguide, with input light end-fire coupled from a SiN waveguide on the same device layer. In contrast to the more common interlayer coupling, such end-fire-coupled devices are a novel addition to SiN photonics platforms.

We fabricate devices with different device geometries and doping profiles, and characterize their performance, including dark current, gain, responsivity, dynamic range and bandwidth. We then benchmark our results against other recently reported integrated APDs, and show that our devices are indeed very competitive across multiple performance metrics.

**Results**

*Device Design*

Our device structure is shown in Fig. 1. The primary photodetector structure is a Si rib waveguide of length 16 µm, which has a high absorptivity at visible wavelengths (> 96% at 685 nm). Input light is end-fire coupled from an input SiN rectangular waveguide, which allows for low-loss propagation of visible light [24]. Both the Si APD and SiN input waveguide have the same width $W$. Two values of $W$, 750 nm and 900 nm, are explored. The height of both the Si APD and

SiN waveguide is fixed at 250 nm, and the Si rib height at 125 nm. The structures are fabricated on a SOI substrate on the same device layer, and are cladded with 3 µm of silicon dioxide ($SiO_2$) above and below.

To establish electrical connections to the device, metal electrodes are deposited on top of heavily doped $p^{++}$ and $n^{++}$ regions at the far ends of the Si slab along the *x* axis, 3 µm apart.

A careful consideration of the doping profile is required to produce high-performance APDs. Here, we design our APDs to consist of a p-$n^+$ diode in two different doping configurations: lateral and interdigitated (see Figs. 1(b,c)). Both profiles aim to maximize the spatial overlap between the depletion region on the p-doped side and the optical waveguide mode.

The lateral doping profile features a single continuous junction placed asymmetrically along the length of the APD. The design distance between the junction and the $n^+$ edge of the waveguide ∆j is {120, 150} nm for waveguide widths $W$ = {750, 900} nm. We have previously performed simulation studies of this doping profile in Si APDs [25, 26]. Though conceptually simple, this profile requires stringent control of the fabrication process, as a small misalignment of the junction will result in a large mismatch between the optical mode and the depletion region.

The alternative design uses an interdigitated profile, which consists of alternating p and $n^+$ regions, each 1 µm in length. This design is less sensitive to such misalignment errors, but the increased junction lengths could lead to a higher depletion capacitance and hence limit the bandwidth, as is reported for Si modulators [27, 28].

The $n^+$ (p) doping concentrations of $1 \times 10^{19}$ ($2 \times 10^{17}$) dopants/cm$^3$ are chosen to ensure that the depletion region covers a large part of the waveguide width in both doping profiles. These doping concentrations are similar to values in other APDs [29–31].

In most recent reports on waveguide-based APDs for infrared wavelengths, input light is coupled to the detector via a phase-matched interlayer transition [21–23]. However, this is challenging to achieve in a SiN ($n$ = 2.1) to Si ($n$ = 3.8) transition due to the large difference in refractive indices. Therefore, we choose to end-fire couple the input SiN waveguide to the Si rib waveguide in the same layer. From our previous analysis of the optical mode overlap between the waveguide modes, we expect a SiN-Si end-fire coupling loss of ≤ 1 dB per facet [25].

In our fabricated devices, light is coupled into the SiN waveguides via inverse tapers at the edge of the waveguide chip (see Fig. 1(e)). For both waveguide widths, the inverse tapers are designed to have a taper length of 200 µm and a minimum taper width of 180 nm. The edge-coupled devices are optimized for interfacing with lensed optical fibers; for a focused spot diameter of 2 µm, the expected coupling loss into the SiN waveguide is ∼ 1.5 dB per facet. Detailed characterization of the coupling and propagation losses yield a total insertion loss of 7.1 ± 0.4 dB for our devices (see paragraph S1 of Supplementary Information).

*Current-voltage measurements*

We measure the current-voltage (I-V) characteristics of each device up to the breakdown voltage $V_{br}$, with a series of different input optical powers $P_{opt}$ entering the Si waveguide. The values of $P_{opt}$ are reported after accounting for the insertion loss. Here we consider representative results for a $W$ = 900 nm laterally doped device, as shown in Fig. 2(a).

From the I-V data we extract two important performance metrics: the responsivity $R$ and the avalanche gain $G$ (Fig. 2(b,c)). From the photocurrent $I_{ph} = I_{dev} - I_{dark}$, where $I_{dev}$ is the measured device current, we obtain the responsivity $R = I_{ph}/P_{opt}$. The avalanche gain $G$ is defined here as the ratio of the photocurrent $I_{ph}$ at bias $V_B$ to that measured at a low bias regime of 1 V [10], where the effects of avalanche gain are negligible:

$$G = \frac{I_{ph}(V_B)}{I_{ph}(1\,\text{V})} \qquad (1)$$

At $V_B$ > 10 V, both $I_{dark}$ and $G$ increase dramatically due to avalanche multiplication. In this regime, the power dependence of the device response becomes obvious, with $G$ and $R$ decreasing for higher $P_{opt}$. This is due to the larger number of multiplied charge carriers causing an increased space charge effect. As a result, the electric field is depressed, leading to saturation of the device current. Thus, while $G \sim 10$ at $V_B = 15$ V for $P_{opt} = $ -20 dBm, it rises to $G > 10^5$ for a low input power of $P_{opt} = $ -71 dBm. Power-dependent characteristics have also been studied in other APDs [32, 33].

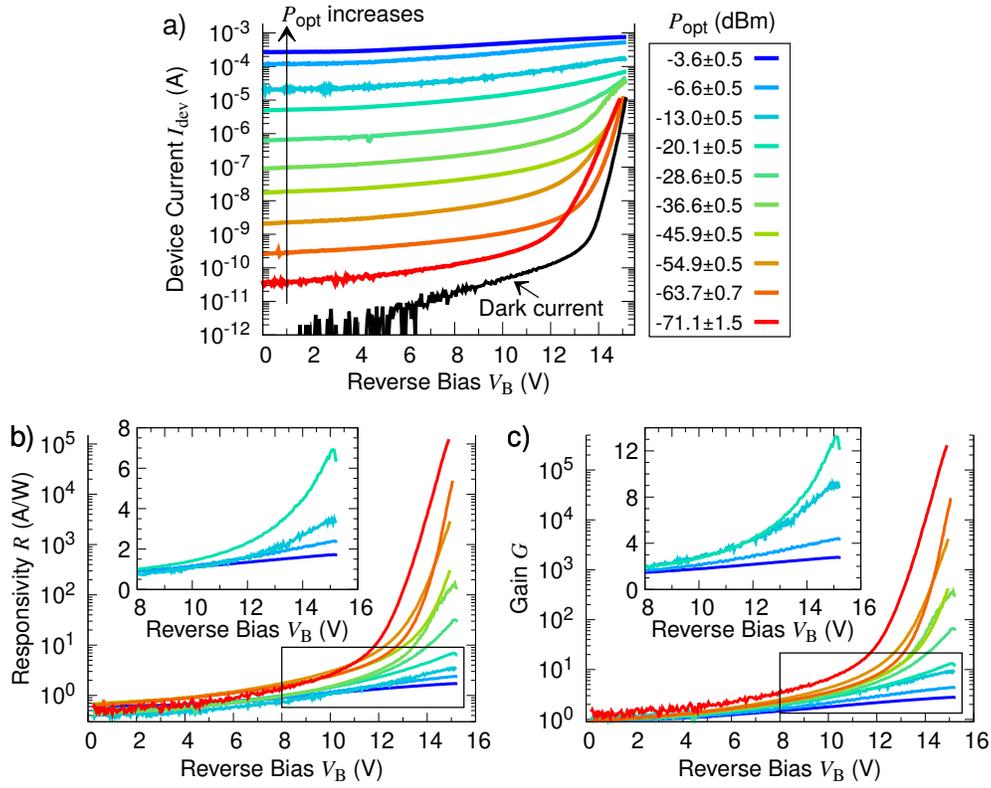

Fig. 2. DC characteristics of a laterally doped device with width $W = 900$ nm. (a) Current-voltage measurements at different input optical powers $P_{opt}$. The reverse bias voltage $V_B$ is swept till the avalanche breakdown voltage $V_{br} \approx 15.5$ V, where the dark current $I_{dark}$ reaches 10 μA. Each sweep takes a few seconds; prior to each sweep, the device is reset with the application of a forward bias voltage. (b),(c) The responsivity $R$ and avalanche gain $G$ at different $P_{opt}$, respectively. The inset is a magnified view of the area marked by the rectangle, showing the curves at larger $P_{opt}$ on a linear scale. All plots in this Figure share the same legend for $P_{opt}$.

As such, we will separately compare the device performance in low-gain and high-gain regimes.

### Performance in the low-gain regime

In the low-gain regime, the APDs can be operated at small bias voltages suited for applications requiring low power consumption. An important example is to monitor optical power levels in integrated photonic circuits, which requires low dark current and wide dynamic range with linear response [10, 32].

We focus on the primary responsivity $R_p$ measured at $V_B = 1$ V, where the gain $G \sim 1$: all device types show linear behavior, with $R_p$ within an overall range of $0.51 \pm 0.11$ A/W over a dynamic range of > 50 dB (see Fig. 3(a) and Table 1). We expect the actual dynamic range to be even larger since we did not explore higher input powers in detail for all devices, and we had not yet observed the device approaching saturation. $R_p$ is slightly higher for $W = 900$ nm devices due to the larger absorption volume of a wider waveguide.

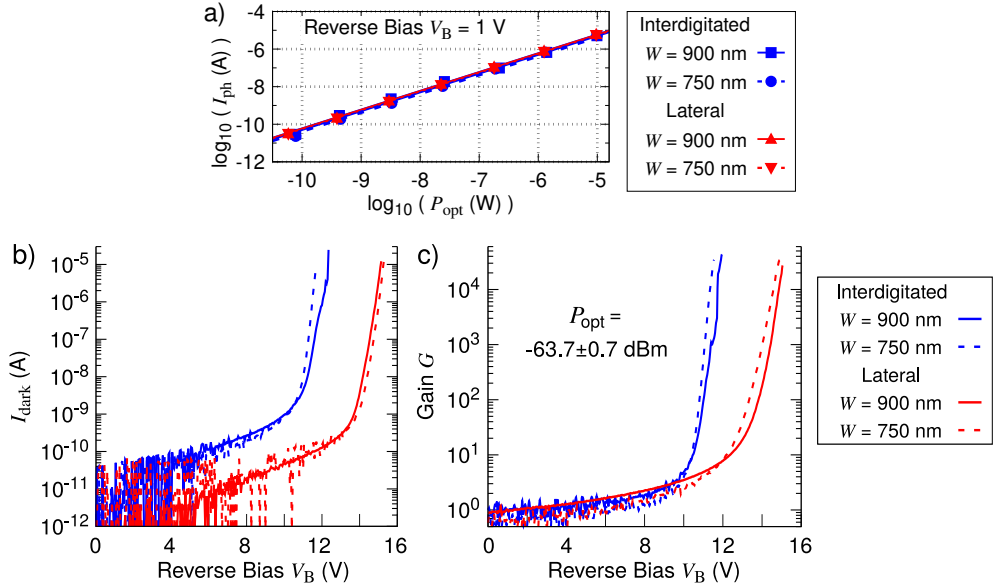

Fig. 3. Comparison of DC performance for lateral and interdigitated doping profiles with different widths $W$. (a) Photocurrent $I_{ph}$ versus input power $P_{opt}$ at reverse bias $V_B = 1$ V, where gain is negligible. Straight lines are linear fits, from which we extract the primary responsivity $R_p$, see Table 1. (b) Dark current $I_{dark}$ measurements at varying $V_B$. (c) Avalanche gain $G$ at varying $V_B$ with a fixed input power $P_{opt} = -63.7 \pm 0.7$ dBm. Figures (b) and (c) share the same legend on the right.

The dark current measurements are shown in Fig. 3(b). $I_{dark}$ at $V_B = 1$ V is less than 50 pA for all device types. Laterally doped devices with $W = 900$ nm exhibit the lowest $I_{dark} < 1$ pA (see also Fig. 2(a)). We note that in the low bias regime ($V_B < 10$ V), laterally doped devices have about an order of magnitude lower $I_{dark}$ than interdigitated devices. This effect has also been previously reported in other waveguide-based photodetectors [32]. There are two likely reasons for the higher dark current in interdigitated devices. First, high peak electric field strengths associated with the corners of the interdigitated regions can lead to a higher dark carrier generation rate [26] (see paragraph S3 of Supplementary Information for more details on the electric field profiles). Furthermore, the interdigitated devices have a larger depletion volume where dark carriers can undergo avalanche multiplication, compared to their laterally doped counterparts.

*Performance in the high-gain regime*

Fig. 3(c) shows the gain $G$ for different device types at a relatively low input power of $P_{opt} = -63.7 \pm 0.7$ dBm, where the devices exhibit high gain. We see that interdigitated devices have a lower breakdown voltage $V_{br}$ and a slightly steeper rise in $G$ with respect to $V_B$. These effects can likely be attributed to premature breakdown due to high electric fields at the edges of the interdigitated regions. For both doping profiles, we observe no significant dependence of $V_{br}$ on the device width $W$. This is consistent with our previous simulations for laterally doped devices [25].

Applications in integrated photonics typically require low power consumption, thus both $I_{dark}$ and $V_B$ should ideally be low as well [10]. While interdigitated devices achieve similar gain at a lower $V_B$ compared to laterally doped devices, $I_{dark}$ tends to be higher. The optimal choice of doping profile in this regime would then require a more in-depth consideration of the operating requirements.

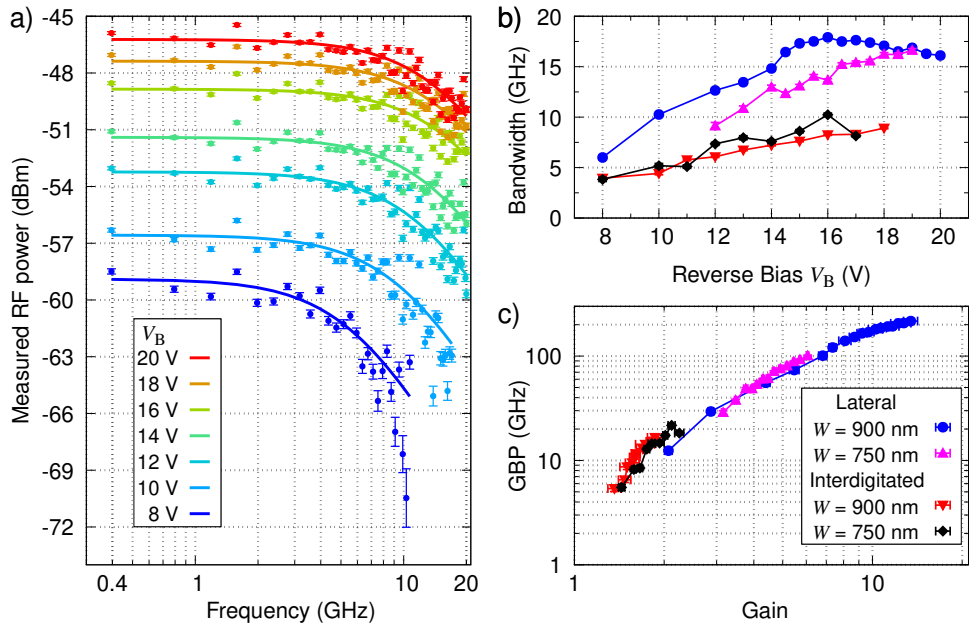

Fig. 4. Optical-electrical bandwidth measurements. Input powers are $P_{opt} = -22.6 \pm 0.7$ dBm and $-17.6 \pm 0.7$ dBm for devices with lateral and interdigitated doping profiles, respectively. (a) Frequency response of a $W = 900$ nm laterally doped device at various bias voltages $V_B$. The bandwidth is obtained from the 3 dB roll-off frequency of a fit to the transfer function of a first order low pass filter. The main contributors to the error bars are sensor noise and impedance mismatch. (b),(c) 3 dB bandwidth and gain-bandwidth product (GBP), respectively, for different device types. Both plots share the same legend shown in (c).

*Frequency Response and Bandwidth*

The AC response of the APDs is characterized in the low-gain operation mode, after the device gain has stabilized (see Methods). Fig. 4(a) shows the results of a frequency response measurement for a $W = 900$ nm laterally doped device. The device bandwidth is obtained from the 3 dB roll-off frequency from a fit to the transfer function of a first order low pass filter. The uncertainty in the measurements are largely due to sensor noise and impedance mismatch.

Figs. 4(b,c) compare the bandwidth and gain-bandwidth product (GBP) of the different device types. In general, the bandwidth increases with higher reverse bias $V_B$ due to a wider depletion region and a lower junction capacitance. However, this effect eventually reaches a limit, beyond which the bandwidth does not increase further with $V_B$; this is clearly visible in laterally doped devices.

We find that the bandwidth is indeed lower in interdigitated devices, as expected from the higher capacitance due to its doping profile. Another potential contributing factor is that a larger proportion of photo-generated charge carriers in interdigitated devices are created in n$^+$-doped regions where the electric field is low, leading to slower carrier diffusion and hence slower device response (see paragraph S3 of Supplementary Information).

Table 1. Benchmarking of device performance with other recent reports of integrated APDs. Results from this work are indicated in blue.

| Type | $\lambda$ (nm) | $V_B$ (V) | $I_{dark}$ (µA) | $R_p$ (A/W) | Gain | BW (GHz) | GBP (GHz) | Device/ Ref. |
|---|---|---|---|---|---|---|---|---|
| Si, LD | 685 | 20 | 0.12(1) | 0.60(2) | 13.4(7) | 16.1(1) | 216(12) | $W$ = 900 nm |
| Si, LD | 685 | 19 | 0.037(7) | 0.55(1) | 6.06(6) | 16.6(2) | 101(2) | $W$ = 750 nm |
| Si, ID | 685 | 18 | 0.31(4) | 0.55(5) | 1.9(1) | 8.91(6) | 16.6(9) | $W$ = 900 nm |
| Si, ID | 685 | 16 | 0.034(3) | 0.43(3) | 2.12(6) | 10.2(2) | 21.7(7) | $W$ = 750 nm |
| Si | 850 | 14 | 2 | 0.05 [a] | 6 | 16.4 | 98.4 [a] | [9] |
| Si | 850 | 20 | 0.075 | 0.2 | 1.45 [a] | 14 | 20.3 [a] | [11] |
| InAs | 1310 | 18.6 | 2000 | 0.06 [a] | 45 | 5.3 [a] | 240 | [4] |
| InAs | 1310 | 15.9 | 0.0013 | 0.234 | 20.5 [a] | 2.1 | 42.3 [a] | [34] |
| Ge/Si | 1310 | 12 | 100 | 0.65 | 11 | 27 | 300 | [35] |
| Ge/Si | 1310 | 11 | 10 | 0.6 | 1.8 | 56 | 101 [a] | [36] |
| Ge/Si | 1310 | 18 | 270 | 0.6 | 10 | 36 | 360 [a] | [36] |
| Si | 1550 | 9 | 88000 | 0.0005 [a] | 1080 [a] | 26 | 28000 | [10] |
| Ge/Si | 1550 | 6 | 1000 | 0.48 | 15 | 18.9 | 284 [a] | [7] |
| Ge/Si | 1550 | 10 | 1 | 1.25 | 17.8 | 25 | 445 [a] | [37] |

[a] These values were not directly reported, and are inferred from other values in these references.

LD: lateral doping  ID: interdigitated doping  $W$: waveguide width
$\lambda$: operating wavelength  $V_B$: reverse bias  $I_{dark}$: dark current
$R_p$: primary responsivity  BW: 3 dB bandwidth  GBP: gain-bandwidth product

A detailed comparison of the best GBP performance for each device is shown in Table 1. The highest observed GBP is $216 \pm 12$ GHz for the $W$ = 900 nm laterally doped device. Although its $W$ = 750 nm version has a lower GBP, it also has a much lower dark current. As such, the optimal choice of device parameters might also depend on the specific application and operating conditions.

The observed trends suggest that an even larger GBP can be obtained if the APD is operated at a higher gain. Unfortunately, we are limited by the wideband 40 GHz RF sensor yielding a thermal noise floor of ∼ -74 dBm. Thus we have to use a relatively high $P_{opt}$ to obtain a signal appreciably above the noise floor, which limits the gain at which we can measure the bandwidth.

We also note that the measured bandwidth can be adversely affected by factors such as the size of contact pads, which could be further reduced or removed altogether in future large-scale integration with a photonics platform.

## Discussion

Table 1 shows the benchmarking of our device performance with other recent reports of integrated APDs. We omit uncertainty values for all the devices we benchmark against, as only some of the literature reports include this information. For each of our devices, we report the performance at the operating conditions where the maximum GBP is observed. We note that the values for the dark current $I_{dark}$ in Table 1 are measured in a different regime compared to Fig. 3(b), where the reset procedure is used.

Our best-performing device is the $W = 900$ nm laterally doped APD, with a GBP of 216 GHz. Compared to other contemporary devices, this APD shows a strong, balanced performance in the performance metrics of dark current $I_{dark}$, primary responsivity $R_p$, gain, and bandwidth. With the exception of Ref. [10] which has a very high operating $I_{dark}$ of 88 mA, the 216 GHz GBP of our APD is also comparable to the highest reported values of a few hundred GHz. Yet, our APD also exhibits a much lower $I_{dark}$ of 0.12 µA at the operating bias $V_B$ than other high-GBP devices; this would lead to decreased noise and power consumption.

These observations show that our devices are competitive, and are well-suited for visible-light applications requiring high bandwidth and high sensitivity.

## Conclusion

We have reported the first fabrication and characterization of waveguide-integrated Si APDs for visible light (685 nm). Our devices feature a small device footprint and are fabricated with a CMOS-compatible process. At a reverse bias of $V_B = 1$ V, a laterally doped APD of 900 nm width exhibited a highest primary responsivity of $0.60 \pm 0.02$ A/W over a dynamic range of > 50 dB, with dark current < 1 pA. At higher $V_B$, laterally doped devices exhibit superior bandwidth, with a highest gain-bandwidth product of $216 \pm 12$ GHz. APDs with an interdigitated doping profile require a lower bias to attain the same DC gain than lateral ones, but have a higher dark current. Our devices perform strongly compared to other state-of-the-art integrated APDs operating at other wavelengths.

The addition of integrated visible-light APDs to the component toolbox of SiN photonics opens up many application possibilities, and greatly expands the versatility of silicon photonics platforms [38]. There is potential for further design optimizations, such as an alternative doping profiles [10] which may enhance the APD gain and reduce the working bias. Future work will also explore the operation of these devices in the Geiger mode for single-photon counting, which will play an important role in the development of integrated quantum photonics platforms, and for interfacing with single-photon sources operating at visible wavelengths.

## Methods

### Device Fabrication

The devices were fabricated in a CMOS foundry. The main fabrication steps of the device are as follows: we start from an 8-inch SOI wafer, with 220 nm Si and 3 µm buried oxide (BOX) layers. An epitaxy of Si (30 nm) tops up the total Si thickness to 250 nm. We then form the Si slab using 248 nm KrF deep-UV lithography and inductively coupled plasma (ICP) etch.

We then deposit a 450 nm-thick SiN layer using low pressure chemical vapor deposition (LPCVD), and reduce it to the same height as the Si slab (250 nm) with chemico-mechanical polishing (CMP) followed by wet etch. Next, we use lithography and ICP etch with an oxide hard mask to pattern the SiN waveguides and then the Si rib waveguides (125 nm etch into the Si slab) in subsequent steps. The image shown in Fig. 1(d) was taken after removing the oxide hard mask.

We perform implantation of the p and n$^+$ regions along the Si rib waveguide, followed by the p$^{++}$ and n$^{++}$ ohmic contact regions with a subsequent rapid thermal anneal at 1030 °C for 5 s. We then deposit 3 µm of oxide as the top cladding, followed by the opening of contact holes. Finally, we deposit and pattern aluminium to form the contact pads.

*Characterization setup*

We test the fabricated devices at room temperature using a custom-built light-tight probe station (see paragraph S2 of Supplementary Information for the setup schematic). We establish electrical connections via $100 \times 100$ µm contact pads on the chip surface using electrical probes (see Fig. 1(f)). We use a 685 nm continuous wave diode laser as the optical source. The laser is coupled to the SiN waveguide using single-mode tapered lensed fibers (OZ Optics, 2 µm spot diameter).

We maintain a horizontal input polarization, which couples to the fundamental TE mode of the SiN waveguide. Although different input polarizations could lead to some variations in the coupling and propagation losses, the APD response itself is not expected to exhibit any significant polarization dependence.

*Electro-optic characterization*

For I-V measurements, the reverse bias voltage $V_B$ is swept from 0 V to the avalanche breakdown voltage $V_{br}$ over a few seconds. We define $V_{br}$ here as the voltage where the dark current $I_{dark}$ (i.e. without input light) reaches 10 µA; this definition follows other reports of APDs in the literature [39, 40]. We note here that $V_{br}$ drifts with time in our devices; as such, to ensure consistent results, it is necessary to reset the device with the application of a forward bias voltage prior to each sweep. More details regarding the drift behavior is discussed in paragraphs S4 and S5 of the Supplementary Information.

For bandwidth measurements, the device gain is first stabilized by continuously applying a reverse bias over $\sim$ 30 mins; this is necessary due to the drift behavior. The 685 nm input light is modulated with an RF signal using an electro-optic modulator (EOM). The EOM is maintained at its half transmission point, i.e. the DC bias is adjusted such that the EOM output power is at 50% of its maximum value, before RF modulation is added. The modulation frequency is spanned from 400 MHz up to 20 GHz. The AC component of the photocurrent is measured with an RF power meter (Boonton 51072A sensor, 40 GHz bandwidth) through a bias tee, and corrected for the frequency-dependent losses in the setup, which are independently measured. For laterally doped devices, we use an input power of $P_{opt}$ = -22.6 ± 0.7 dBm. For interdigitated devices, which has a lower gain at high frequencies, we increased this to $P_{opt}$ = -17.6 ± 0.7 dBm to ensure a high signal-to-noise ratio.

## Funding

We acknowledge the support of the National Research Foundation, Singapore (NRF-CRP14-2014-04).

## Acknowledgments

S. Y. would like to acknowledge the support of the Singapore International Graduate Award (SINGA). The authors thank Mingbin Yu and Shiyang Zhu for their inputs at the initial stage of the project.

## Author Contributions

S.Y. and V.L. designed and performed the experiments. J.R.O. led the device design and simulation work. H.T. contributed to the device fabrication. L.K. conceived the idea. C.E.P. and L.K. supervised the work. All co-authors contributed to writing and proofreading the manuscript.

## Disclosures

The authors declare that there are no conflicts of interest related to this article.

# Integrated Avalanche Photodetectors for Visible Light: Supplementary Information

Salih Yanikgonul,[1,4] Victor Leong,[1,†] Jun Rong Ong,[2,*], Ting Hu,[3] Ching Eng Png,[2] and Leonid Krivitsky [1]

[1]*Institute of Materials Research and Engineering, Agency for Science, Technology and Research (A*STAR), 138634 Singapore*
[2]*Institute of High Performance Computing, Agency for Science, Technology and Research (A*STAR), 138632 Singapore*
[3]*Institute of Microelectronics, Agency for Science, Technology and Research (A*STAR), 138634 Singapore*
[4]*School of Electrical and Electronic Engineering, Nanyang Technological University, 639798 Singapore*
[†]*victor_leong@imre.a-star.edu.sg*
[*]*ongjr@ihpc.a-star.edu.sg*

## S1. Coupling and Propagation Loss Measurements

We systematically characterized the coupling and propagation losses on our device by performing a series of cutback measurements with test waveguides. In addition to waveguide widths $W = 750$, 900 nm mentioned in the main paper, here we also investigated $W = 450$, 600 nm.

The optical transmission $T$ through the device was obtained by measuring the input power $P_A$ and output power $P_B$ with a pair of lensed fibers (see Fig. S1(a)). Using SiN cutback waveguides of various lengths $l_{SiN}$ (without the Si rib waveguide), we fitted our results using

$$T = P_B/P_A = \eta_{\text{f-SiN}}^2 \, e^{-(\alpha_{SiN} \, l_{SiN})} \tag{S1}$$

to obtain the fiber-waveguide coupling loss $\eta_{\text{f-SiN}}$ and the SiN waveguide propagation loss coefficient $\alpha_{SiN}$. A representative plot is shown in Fig. S1(b). Following this, we measured another series of devices that also included Si waveguides of various lengths $l_{Si}$; fitting our results to

$$T = P_B/P_A = \eta_{\text{f-SiN}}^2 \, \eta_{\text{SiN-Si}}^2 \, e^{-(\alpha_{SiN} \, l_{SiN})} \, e^{-(\alpha_{Si} \, l_{Si})} \tag{S2}$$

we obtained the SiN-Si end-fire coupling loss $\eta_{\text{SiN-Si}}$ and the Si waveguide propagation loss coefficient $\alpha_{Si}$. The measured coupling and propagation losses are shown in Fig. S1(c). We note that we lack test structures for Si waveguides of width $W = 750$ nm; nonetheless we anticipate that the coupling and propagation losses will not significantly deviate from that of the other widths.

The observed fiber-waveguide coupling losses $\eta_{\text{f-SiN}}$ agree with our expected values. The slight increase in $\eta_{\text{f-SiN}}$ with width $W$ is likely due to the larger inverse taper angle, since the taper length and tip width are kept constant for all widths $W$. The end-fire coupling loss $\eta_{\text{SiN-Si}}$ is ∼3-4 dB larger than mode-matching calculations, which is attributed to fabrication imperfections resulting in a non-ideal waveguide interface. We also observed decreasing propagation losses with increasing width $W$.

For APD characterization, the total insertion loss into the active device structure (i.e. the Si rib waveguide) is given by

$$\eta_{\text{total}} = \eta_{\text{f-SiN}} \, \eta_{\text{SiN-Si}} \, e^{-(\alpha_{SiN} \, l_{SiN})} \tag{S3}$$

where $l_{SiN} = 0.3125$ cm is constant for all characterized devices. For both $W = 750$ nm and 900 nm, this yields $\eta_{\text{total}} = 7.1 \pm 0.4$ dB. We decided to focus on devices with lower insertion loss, and thus only considered devices with these two widths for further characterization in the main paper.

## S2. Electro-optic characteriazation setup

The schematic of the electro-optic characterization setup is shown in Fig. S2. The detailed description is found in the Methods section of the main text.

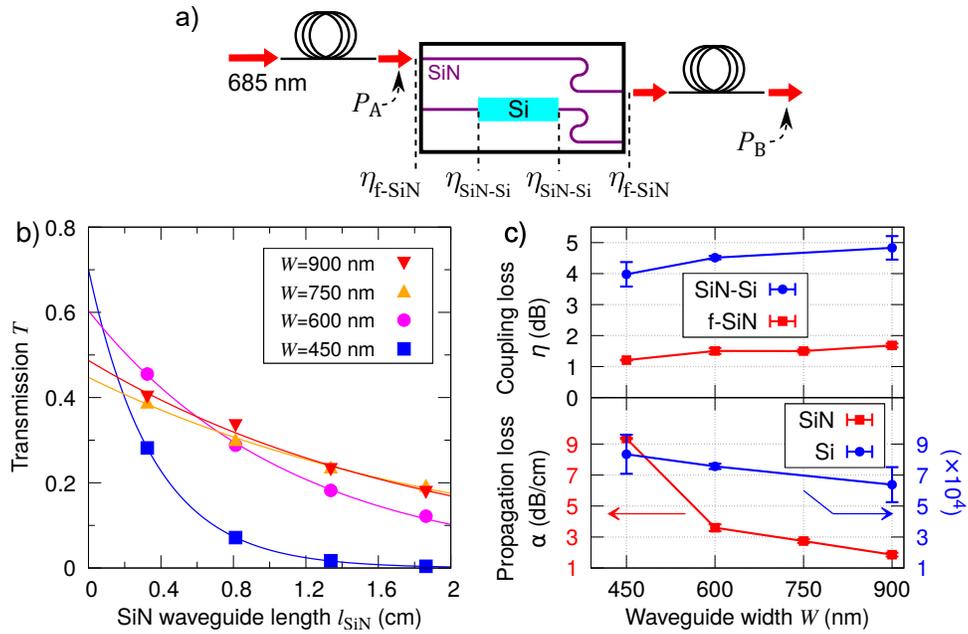

Fig. S1. Optical coupling loss measurements. (a) Schematic of the experimental setup, depicting the cutback waveguide structures and the various sources of coupling losses. Horizontally polarized 685 nm light is coupled to and from the waveguides via lensed fibers. The optical powers at both ends of the chip (denoted $P_A$ and $P_B$) are measured. (b) Optical transmission measurements for SiN cutback waveguides of various lengths $l_{SiN}$ and different waveguide widths $W$. The solid curves are exponential fits, see Eq. S1. (c) Measured coupling losses for different waveguide widths $W$. The results shown in (c) are the averaged measurements across several devices.

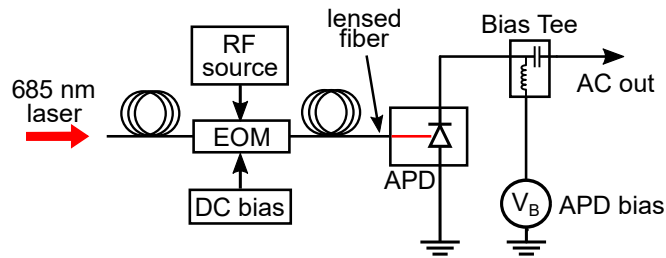

Fig. S2. Schematic of the characterization setup. Horizontally polarized (TE) 685 nm light, which can be modulated with a RF signal using an electro-optic modulator (EOM), is coupled to the on-chip SiN waveguide with a lensed fiber. Electrical connections to the devices are made via contact pads on the chip surface using electrical probes. A bias tee separates the AC and DC signals from the APDs.

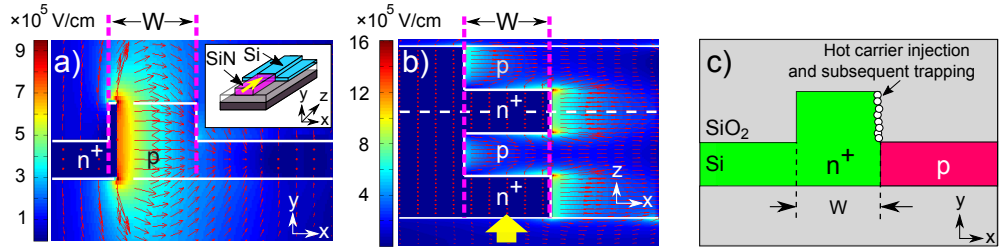

Fig. S3. (a) Electric field profile of a laterally doped device. Inset shows the schematic of the APD structure (top cladding omitted for clarity) and the axis orientations. (b) Electric field profile of an interdigitated device. The highest electric field strengths are concentrated at the corners of the $n^+$-doped areas. The devices in both (a) and (b) have the same width $W$, and are simulated at just above the breakdown voltage. The yellow arrows indicate the propagation direction of input light. (c) Schematic of the waveguide cross-section along the horizontal white dashed line in (b). We mark the interface where we expect significant injection and subsequent trapping of hot carriers.

## S3. Analysis of simulated electric field profiles

In this section, we will analyse the representative electric field profiles of both laterally doped and interdigitated devices of the same width $W$, and relate their features to the device characteristics we observe in our measurements. The electric field profiles are obtained using the ATLAS device simulator (Silvaco Inc).

Fig. S3(a) shows the electric field profile in a laterally doped device, where the high-field regions are found along the p-$n^+$ junction within the waveguide core. More detailed simulation results of laterally doped devices can be found in our previous works [1, 2]. Fig. S3(b) shows the electric field profile of an interdigitated device. To reduce computation time, we limited the scale of the device to only two periods of alternating p-$n^+$ regions. Nonetheless, we are still able to obtain the necessary features for our analysis. Both figures are simulated at just above the breakdown voltage of the devices.

### S3.1. Peak electric field strengths

We observe that the highest electric field strengths in the interdigitated device are concentrated at the corners of the $n^+$-doped areas, and that their magnitude is significantly higher than that found in the laterally doped device with the same waveguide dimensions. The emergence of these localised high-field regions is likely to have resulted in a lower breakdown voltage $V_{br}$ in interdigitated devices. This could also have contributed to the higher dark current observed in interdigitated devices, due to the exponential dependence of the dark carrier generation rate on higher field strengths.

### S3.2. Light absorption in undepleted $n^+$-doped regions

In a p-$n^+$ junction, the p-doped regions are fully depleted, but the depletion region only extends minimally into the $n^+$-doped regions due to their higher doping concentration. For the lateral doping profile, there is a large overlap between the depletion region and optical mode over the full length of the Si rib waveguide. However, for the interdigitated design, a significant amount of light absorption occurs in the undepleted $n^+$-doped regions, as input light propagates along the alternating p- and $n^+$-doped "digits".

It is less desirable for light absorption to occur in the undepleted $n^+$ regions. Due to the weak electric field strengths, the avalanche multiplcation of the photo-generated charge carriers is less efficient compared to the high-field depletion region, and thus it is detrimental to the overall device gain and responsivity. This might explain the slightly lower responsivity observed in $W = 750$ nm interdigitated devices compared to laterally doped devices, though we do not observe a significant difference for $W = 900$ nm devices. The slower charge carrier diffusion in the low-field regions [1] would also contribute to the lower device bandwidth observed in our interdigitated devices.

In our devices, input light is first incident on a $n^+$-doped region. This results in an overall slightly higher absorption (a difference of ~10%) in $n^+$-doped compared to p-doped regions. Thus, the effect of light absorption in the $n^+$-doped regions could be slightly reduced by having input light incident on the opposite end of the Si waveguide, such that the light is first incident on a p-doped region.

The dimensions of the interdigitated doping regions can potentially be optimised, e.g. the pitch and length of each doping region, or to have p- and $n^+$-doped regions of different lengths. However, we foresee a design trade-off as increasing the depletion volume would also likely increase the device capacitance, which could lead to an RC-limited bandwidth.

### S3.3. Charge trapping

Charge trapping can occur as hot carriers are injected into the $SiO_2$ cladding, and are subsequently trapped at the interface between the $n^+$-doped region and the cladding. The trapped charges would change the electrical field distribution inside the depletion region over time [3]. This effect is likely more severe in interdigitated devices, as the high electric fields occur at the edge of the Si rib waveguide (see Fig. S3(c)). This could lead to drifts in breakdown voltage and device gain over time, which is discussed in detail in the next section.

A potential mitigating strategy is to include guard-ring structures [4] at the Si-$SiO_2$ interface. In addition, adopting a shallow etch for the Si rib waveguide would also reduces the interface area for charge trapping.

### S4. Decaying gain and breakdown voltage drifts at high bias voltages

In our devices, the device breakdown voltage $V_{br}$ drifts towards higher values over time as a reverse bias voltage $V_B$ is continuously applied. This is accompanied by an observed decay in the photocurrent and dark current from the onset of applying the reverse bias. Representative measurements based on a $W = 900$ nm laterally doped device are shown in Fig. S4(a),(b). The decrease is more pronounced at higher $V_B$, with a steep decay in the current at the start before gradually leveling off; while at lower $V_B$ the decay is much slower.

There is a corresponding decrease in the gain $G$ with time, as shown in Fig. S4(c). While the effect is minimal at lower $V_B$, where $G$ decreases by <10% over 10 mins for $V_B <13$ V at $P_{opt} = -25$ dBm, the drop in gain increases sharply at high $V_B$. The rate of decrease slows down significantly after the first 10 mins, but full stability of $G$ is observed only after ~ 30 mins.

The $V_{br}$ drift has been reported in other APDs [5,6]. As discussed above, this effect is likely to be more severe in interdigitated devices. This is consistent with our observation of a larger gain reduction over time for interdigitated devices (comparing Table 1 and Fig. 4 of the main text).

This phenomenon reveals two distinct operating modes for our devices: a gated mode where the APD is operated at high $V_B$ with high gain, using a reset procedure to circumvent the decay in gain (explained in the following section); and a continuous mode where the APD is either operated at low $V_B$, or after the gain has stabilized over some time under a higher $V_B$.

### S5. Reset procedure for gated operation with high gain

Despite the APD gain decreasing over time, it can be reset by the application of a forward bias voltage $V_F$. This likely causes the de-trapping of the charge carriers, allowing the device gain to recover to its original value. The procedure is illustrated in Fig. S5(a). In between the sweeps of the reverse bias $V_B$ used to characterize the APD, we apply $V_F = -1$ V for 1 s to the device cathode; a shorter duration might be sufficient, but we did not investigate this in detail. We note that our measurement instrument limitations result in a delay of ~1 s when switching between $V_B$ and $V_F$.

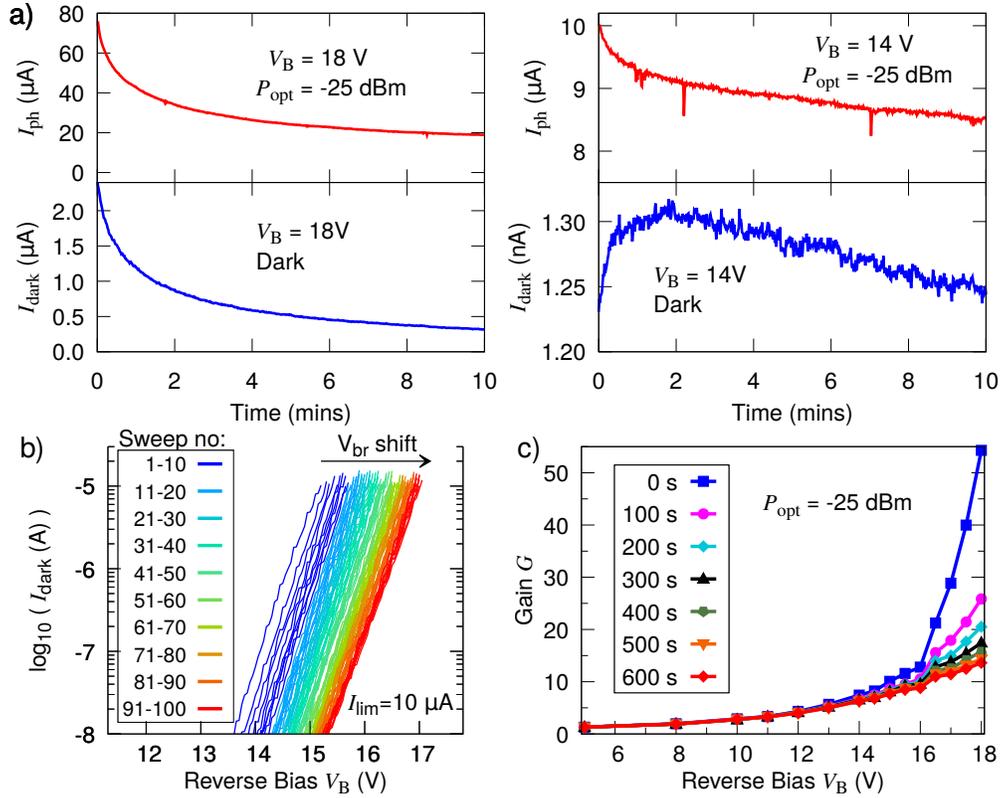

Fig. S4. Decaying gain and breakdown voltage drifts for a $W = 900$ nm laterally doped device. (a) Photocurrent $I_{ph}$ (measured at $P_{opt} = -25$ dBm) and dark current $I_{dark}$ at reverse bias $V_B$ of 14 V and 18 V. Prior to each measurement, the device is reset with the application of a forward bias voltage. (b) The avalanche breakdown voltage $V_{br}$ increases upon successive voltage sweeps. Each sweep starts from $V_B = 0$ V and is terminated upon the dark current $I_{dark}$ reaching the breakdown current of 10 µA. Here, the device is not reset with a forward bias voltage in between runs. (c) Change in gain $G$ over time at different $V_B$. Here, the reverse bias is continuously applied.

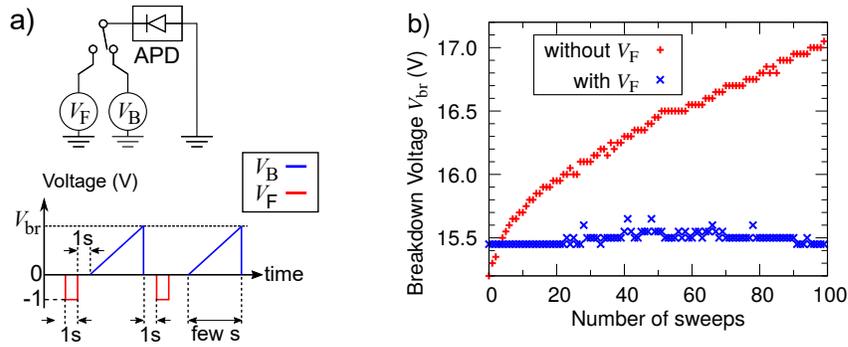

Fig. S5. Reset procedure for high-gain operation. (a) Schematic of the procedure. Here, multiple sweeps of the reverse bias voltage $V_B$ are applied to the APD. Prior to each sweep, a forward bias voltage of $V_F = -1$V is applied to the APD cathode for 1 s. The time axis is not drawn to scale. (b) Comparing breakdown voltage $V_{br}$ drifts for a $W = 900$ nm laterally doped device. By applying the forward bias $V_F$, the breakdown voltage $V_{br}$ remains stable over successive sweeps of $V_B$. If $V_F$ is not applied, we obtain the same results in Fig. S4(b).

The reset procedure prevents the drift in the breakdown voltage over successive voltage sweeps (see Fig. S5(b)), and also results in repeatable current-voltage characteristics after each reset. For the DC characterization results presented in the main report, the reset procedure is carried out before each measurement.

Thus, we are able to periodically operate the APD in the high-gain regime with a gated mode, which would be compatible with applications where gating is used to reduce noise and enhance the signal. Such applications include time-of-flight imaging [7], low-light imaging [8], and Raman spectroscopy [9]. We also note that Geiger-mode infrared InGaAs APDs often employ gating techniques to suppress dark counts and afterpulsing [10].